\documentstyle[psfig,sprocl]{article}

\def\be{\begin{equation}}
\def\ee{\end{equation}}
\def\bea{\begin{eqnarray}}
\def\eea{\end{eqnarray}}

\def\gtrsim{\mathrel{\hbox{\rlap{\hbox{\lower4pt\hbox{$\sim$}}}\hbox{$>$}}}}
\newcommand{\Mpc}{h$^{-1}$Mpc}
\newcommand{\kms}{km s$^{-1}$}

\begin{document}
\title{\bf A SEARCH FOR LARGE-SCALE STRUCTURE AT HIGH REDSHIFT}
\author{A.J. CONNOLLY, A.S. SZALAY}
\address{Department of Physics and Astronomy, The Johns Hopkins
University, Baltimore, MD 21218}
\author{A. K. ROMER, R.C. NICHOL}
\address{Department of Physics, Carnegie Mellon University, 
5000 Forbes Avenue, Pittsburgh, PA 15213, USA}
\author{B. HOLDEN}
\address{Department of Astronomy and Astrophysics, University of Chicago, 
5460 S. Ellis Ave, Chicago, IL 60637}
\author{D. KOO}
\address{University of California Observatories, Lick Observatory, 
University of California Santa Cruz, CA 95064}
\author{T. MIYAJI}
\address{Max-Planck-Institut f\"ur Extraterrestrische Physik
                     Postf. 1603, D-85740, Garching, Germany}
 

\maketitle\abstracts{ We present new and exciting results on our
search for large-scale structure at high redshift. Specifically, we
have just completed a detailed analysis of the area surrounding the
cluster CL0016+16 ($z=0.546$) and have the most compelling evidence
yet that this cluster resides in the middle of a supercluster. From
the distribution of galaxies and clusters we find that the
supercluster appears to be a sheet of galaxies, viewed almost edge-on,
with a radial extent of 31 h$^{-1}$Mpc, transverse dimension of 12
\Mpc, and a thickness of $\sim$ 4 \Mpc. The surface density and
velocity dispersion of this coherent structure are consistent with the
properties of the ``Great Wall'' in the CfA redshift survey. Full
details and followup observations can be found at
http://tarkus.pha.jhu.edu/$\sim$ajc/papers/supercluster/sc.html}

\begin{figure}[t] 
\protect{
\centerline{\psfig{figure=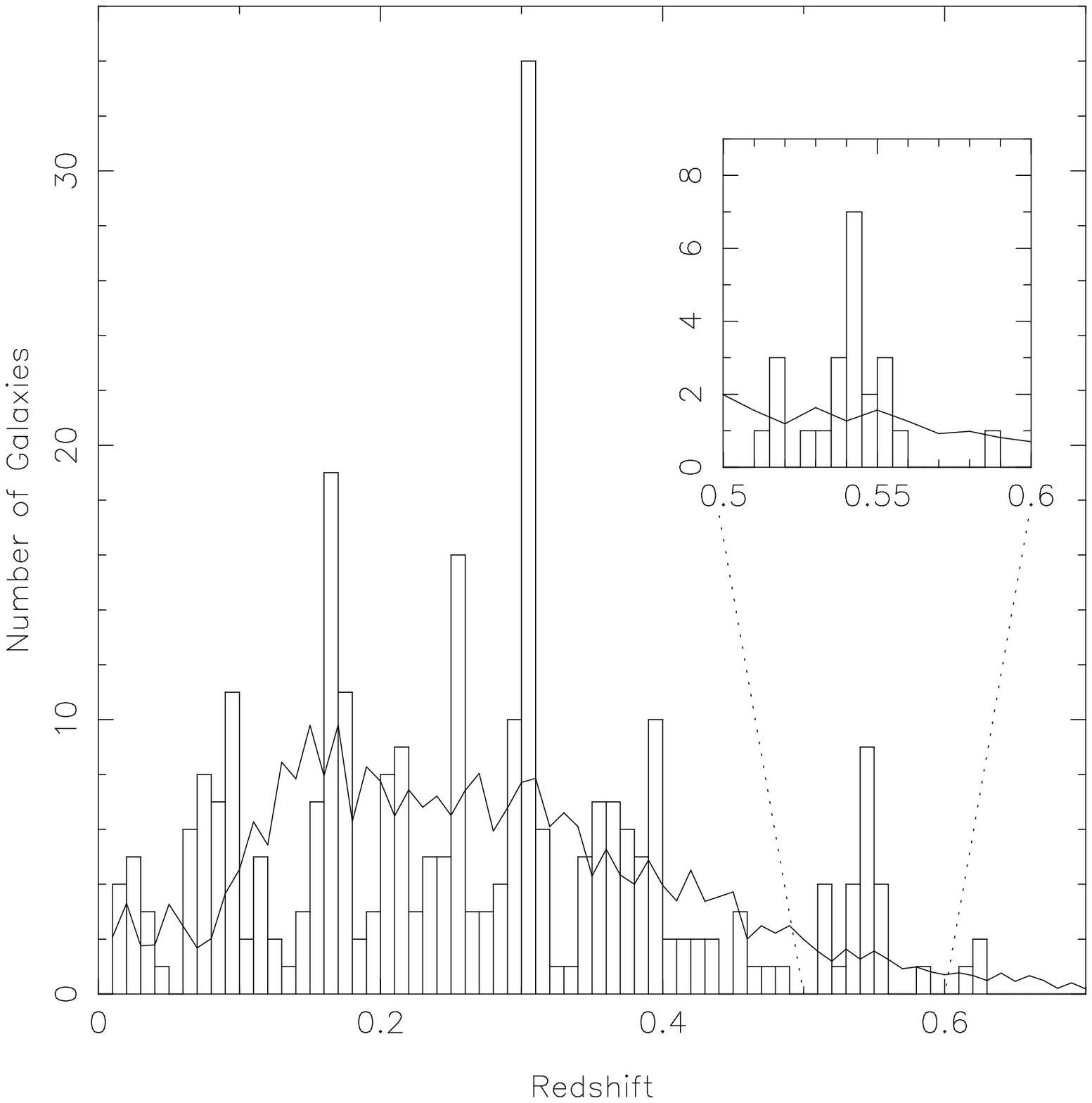,height=2.78in,angle=0.}
\psfig{figure=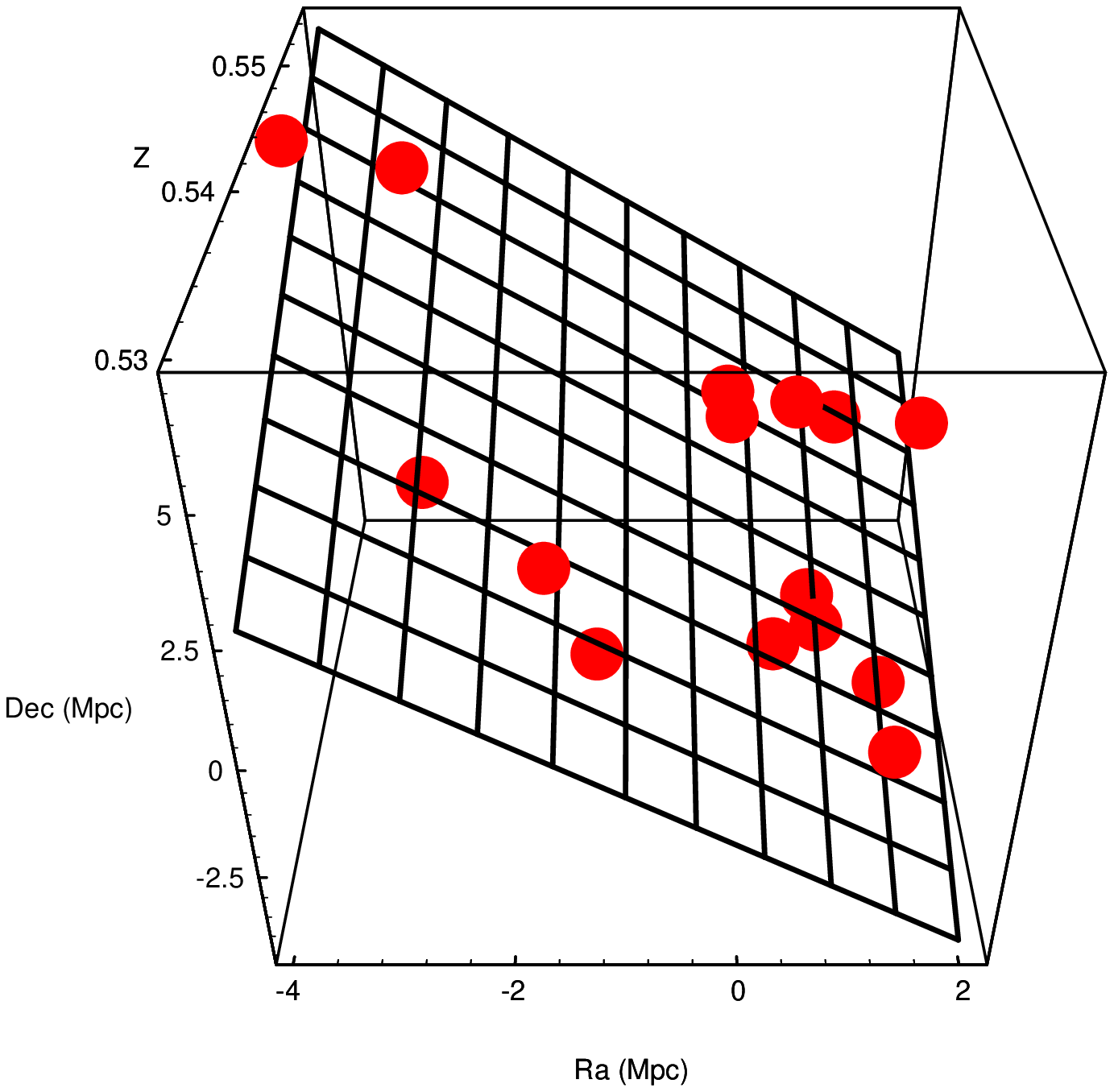,height=2.78in,angle=0.}}}
\caption{(a) The redshift distribution of galaxies in the Selected
Area SA68. The solid line represents the expected distribution of
redshifts for this sample assuming no clustering. The expectation
value for the number of galaxies between $0.530 < z < 0.555$ is $3.54
\pm 3.7$. The observed number of galaxies within this redshift range
is 14, a factor of four larger. (b) Combining the optical and X-ray
data the distribution of galaxies and clusters appears planar. 
The galaxies appear to be part of a 2-dimensional sheet-like
structure which we are viewing edge on (as opposed to a 1-dimensional
filament). The extent of this structure, as defined by the current
data set, is 31 \Mpc\ radially, 13 \Mpc\ in the transverse direction
and with a ``thickness'' of 433 \kms\ orthogonal to the plane.}
\end{figure}

Understanding the clustering of galaxies as a function of redshift
provides important constraints on the initial perturbation spectrum of
the Universe. While in the local Universe extensive redshift surveys
(e.g. CfA) have detected coherent structures of 100 \Mpc\ in extent,
wide angle surveys at $z>0.2$ have been precluded due to the extensive
observational resources required. Consequently, while overdensities in
the redshift distribution of galaxies have been detected their angular
distribution and, therefore, their spatial extent have yet to be well
determined. We consider here the distribution of galaxies in the
direction of Selected Area 68\,\cite{kro80}. We identify this region
as a potential site of an intermediate redshift supercluster because
of a coherent distribution of very red galaxies, estimated to be at $z
\sim 0.5$, and the presence of two nearby X-ray clusters at the same
redshift\,\cite{hug95}.

Figure 1a shows the redshift distribution for those galaxies in SA68
with $B_J <23.0$. The solid line represents the expected redshift
distribution of galaxies assuming an unclustered universe. In the
redshift range $0.530 < z < 0.555$ we find that the observed number of
galaxies exceeds the expected value by a factor of 4 (the expectation
value is $3.54 \pm 3.7$ and the number detected 14). The projected
angular distribution of these galaxies forms a linear structure
passing from the South-West of the SA68 field through to the
North-East.

Along this direction, a little over half a degree from the center of
SA68, lie the X-ray clusters CL0016+16 and J0018.3+1618\,\cite{hug95}
with redshift $z=0.54$. Further, as part of the SHARC survey (see
Romer et al.\ in these proceedings) we have identified an additional
X-ray cluster in the vicinity of CL0016+16\,\cite{con96} (RX
J0018.8+1602). A recent observation with the ARC 3.5m telescope has
confirmed the redshift of this cluster as $z=0.541$ (Figure 2).

\begin{figure}[t] 
\protect{\centerline{\psfig{figure=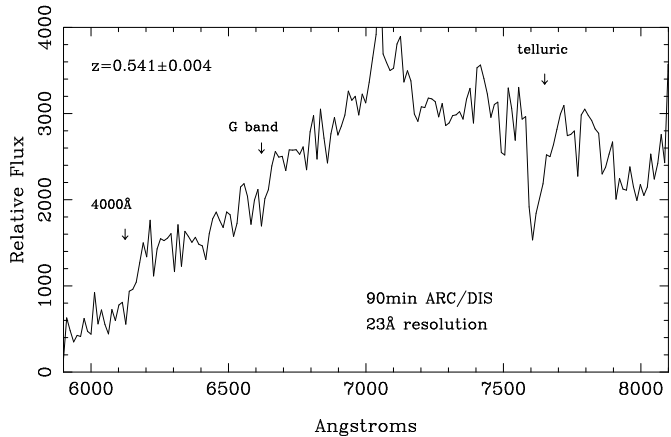,height=2.78in,angle=-90.}}}
\caption{The spectrum for the central elliptical galaxy of
J0018.8+1602 observed with the ARC 3.5m telescope at Apache Point. The
spectrum is consistent with an elliptical at a redshift of
$z=0.541$. }
\end{figure}

Combining the optical and X-ray data we, therefore, have a coherent
structure, at a redshift of $z=0.54$, extending about one degree
across the sky from the survey field SA68 through the cluster
CL0016+16 (see Figure 1b). The positions of the galaxies and clusters
within this volume are not randomly distributed but appear to lie in a
planar distribution (i.e.\ their redshifts and angular distribution
are strongly correlated). Fitting a two dimensional surface to the
spectroscopic redshifts we determine an orientation 40$^\circ \pm
10^\circ$ East of North and an angle 12$^\circ \pm 2^\circ$ from the
line of sight.  The surface density and velocity dispersion of this
supercluster are consistent with the measurements of the ``Great
Wall'' from the CfA survey.


\begin{thebibliography}{99} 
\bibitem{con96} Connolly A.J., Szalay, A.S.,  Koo, D.C., Romer, A.K., Holden,
    B., Nichol, R.C. \& Miyaji, T., 1996 ApJ 473, L67
\bibitem{hug95} Hughes J.P., Birkinshaw, M., Huchra, J.P., 1995, ApJ,
448, L93
\bibitem{kro80} Kron, R.G., 1980, ApJS, 43, 305
\end{thebibliography}
\end{document}